# THE TAC IR FEL OSCILLATOR FACILITY PROJECT*

B. Ketenoglu#, A. Aksoy, O. Yavas, M. Tural, O. Karsli, Ankara University, Ankara, Turkey
S. Ozkorucuklu, S. Demirel University, Isparta, Turkey
P. Arikan, E. Kasap, Gazi University, Ankara, Turkey
H. Yildiz, Dumlupınar University, Kütahya, Turkey
B. Bilen, Doğuş University, İstanbul, Turkey
H. Aksakal, Niğde University, Niğde, Turkey
I. Tapan, Uludağ University, Bursa, Turkey

*Abstract*

The TAC (Turkish Accelerator Center) IR FEL Oscillator facility, which has been supported by Turkish State Planning Organization (SPO) since 2006, will be based on a 15-40 MeV electron linac accompanying two different undulators with 2.5 cm and 9 cm periods in order to obtain IR FEL ranging between 2-250 microns. The electron linac will consist of two sequenced modules, each housing two 9-cell superconducting TESLA cavities for cw operation. It is planned that the TAC IR FEL facility will be completed in 2012 at Gölbasi campus of Ankara University. This facility will give an opportunity to the scientists and industry to use FEL in research and development in Turkey and our region. In this study, the results of optimization studies and present plans about construction process of the facility are presented.

## OVERVIEW OF THE TAC IR FEL OSCILLATOR FACILITY PROPOSAL

The TAC IR FEL project [1] aims to obtain FEL between 2-250 microns range using 15-40 MeV energy range electron beam. In order to have wide research area, it is requested to have cw electron beam with high average current as well as pulsed beam with low current. Therefore, it is planned to use high average current thermionic DC gun and superconducting accelerators with IOT power sources. The electron beam parameters are given in Table 1. To obtain FEL in 2-250 microns range, it was decided to use 2.5 cm and 9 cm period length undulators located in two different optical resonators [2]. The layout of the IR FEL & Bremsstrahlung laboratory is shown in Fig. 1, and the schematic plan of the facility building & Accelerator Technologies Institute (ATI) is shown in Fig. 2.

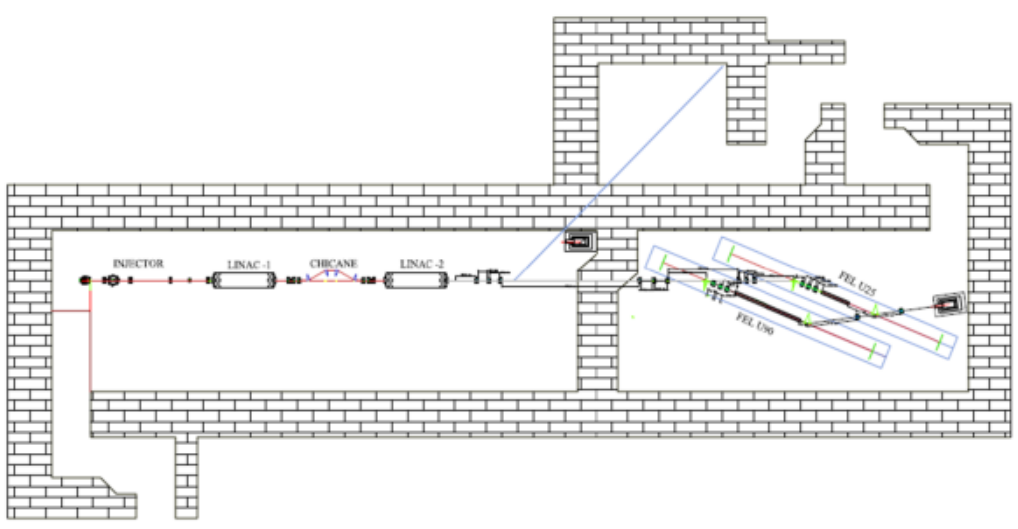

Figure 1: Layout of the TAC IR FEL & Bremss. lab.

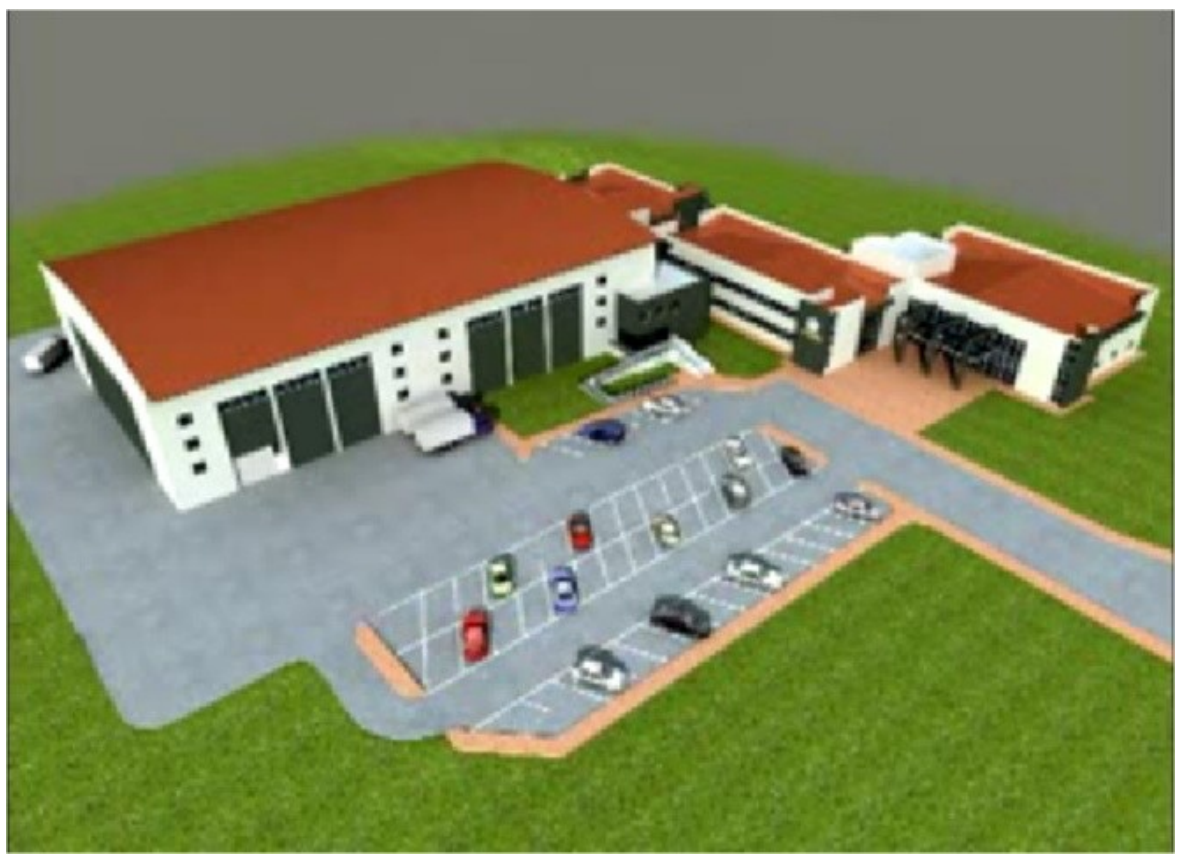

Figure 2: Schematic plan of the facility & ATI @ Ankara

## THE ELECTRON SOURCE & LINAC

An 300 keV high current thermionic DC gun was planned to drive the linac [3]. Schematic view of the injector system is shown in Fig. 3. On the other hand, Fig. 4 represents a detailed view of the gun and the time structure of the electron beam. Furthermore, the electron linac will consist of two sequenced modules, each housing two nine-cell superconducting TESLA cavities for cw operation.

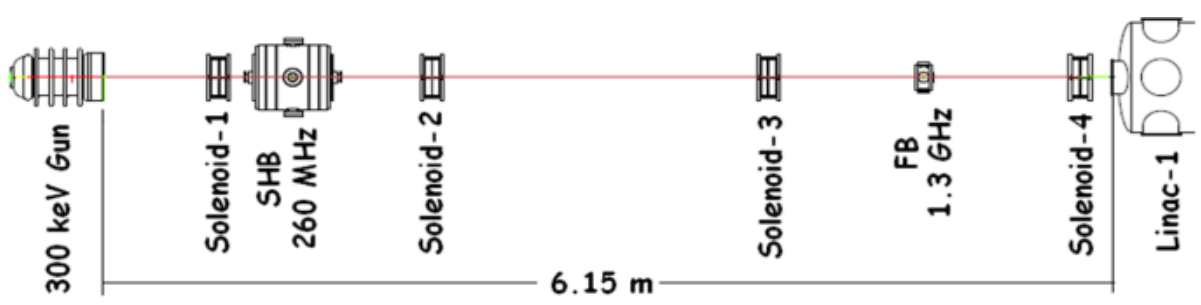

Figure 3: Schematic view of the injector system

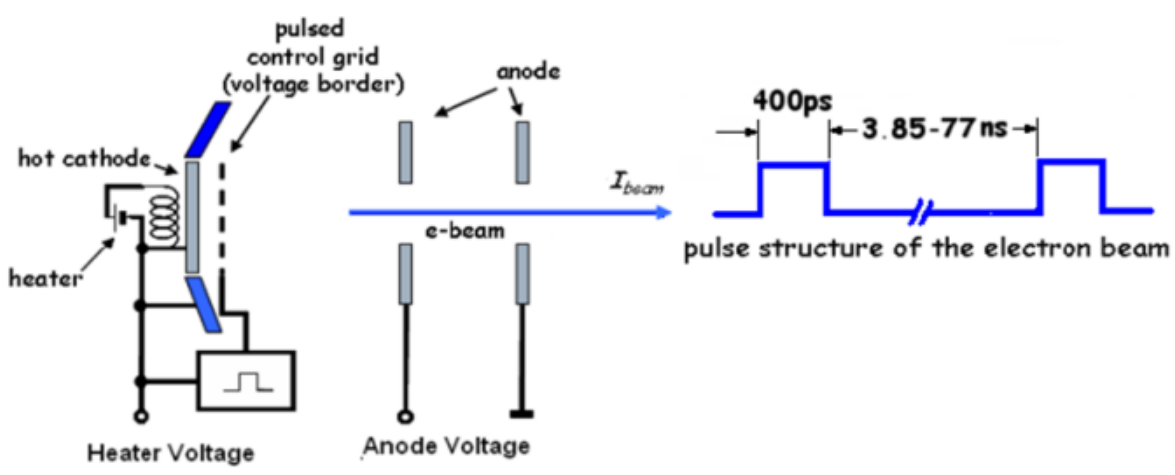

Figure 4: Detailed schematic view of thermionic DC gun & beam time structure

*Work is supported by Turkish State Planning Organization
#bketen@eng.ankara.edu.tr

Table 1: Electron Beam Parameters

| Electron Beam | Value |
|---|---|
| Beam energy (MeV) | 15-40 |
| Average current (mA) | 1.6 |
| Repetition rate (MHz) | 260*- 26 - 13 |
| Bunch length (ps) | 1-10 |
| Norm. rms transerve emittance (mm.mrad) | < 15 |
| Norm. rms longitudinal emittance (keV.ps) | < 100 |
| Macropulse length & repetition | cw / tunable |

*For Bremsstrahlung applications

Research Instruments GmbH offers such accelerating modules with two superconducting TESLA RF cavities for cw operation [4,5]. This module is compact and houses two nine-cell TESLA RF cavities and designed for cw operation at accelerating fields in the range of 10-15 MV/m (see Fig. 5). The cryostat of the module has been developed by ELBE group [6,8]. For cw operation, two SRF ELBE modules were planned to be used in order to achieve the beam energy up to 40 MeV [2]. And a chicane will be located between the modules. In order to obtain 1.6 mA average beam current at 10 MV/m gradient, an 16 kW RF power is required. Tests of ELBE modules are still being continued with higher power than 10 kW. [7]. The schematic view of the main accelerating section is shown in Fig. 6.

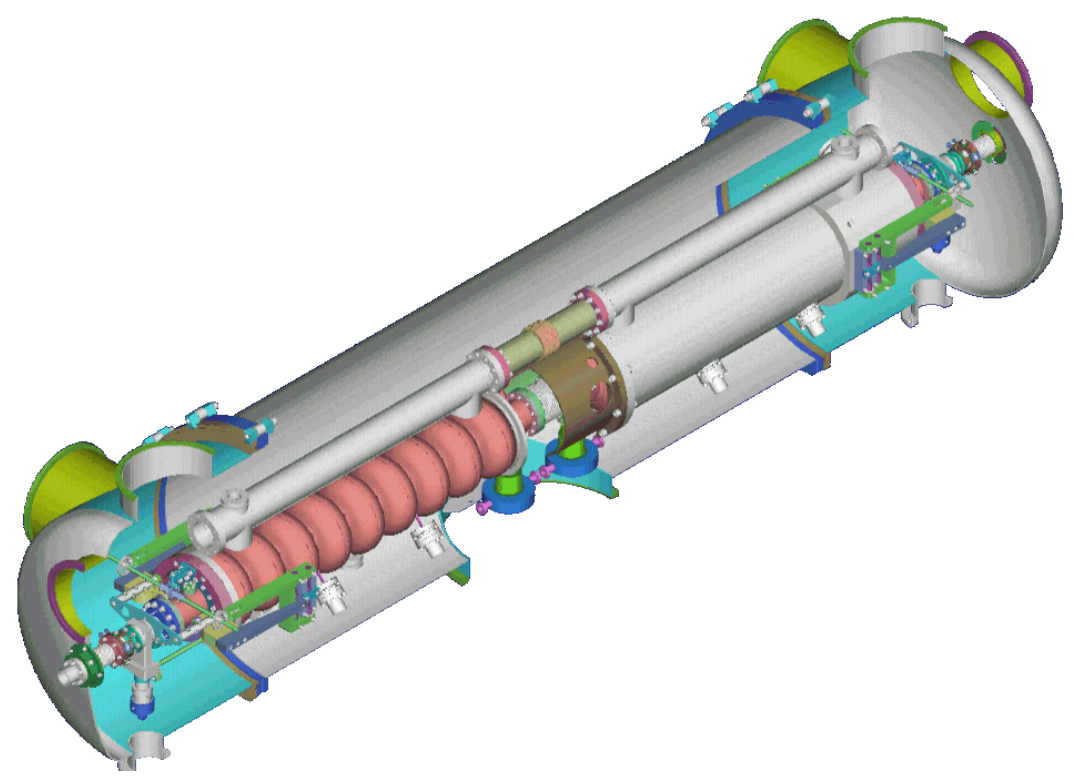

Figure 5: ACCEL (Research Instruments GmbH) module housing two superconducting nine-cell TESLA RF cavities [8].

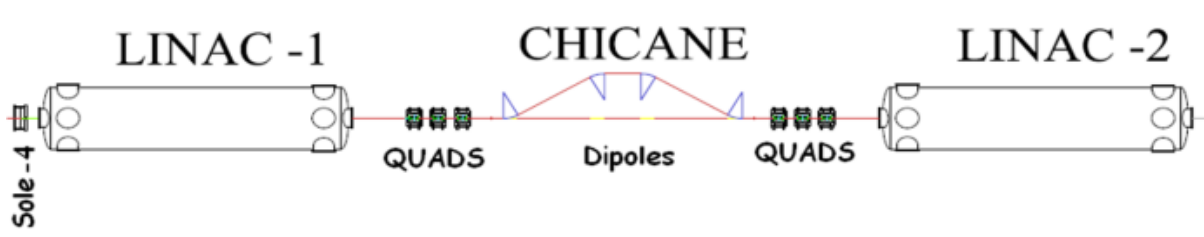

Figure 6: Main accelerating section of the TAC IR FEL oscillator facility.

## THE OPTICAL CAVITIES & UNDULATORS

In order to reduce length of dispersive sections and therefore emittance growth due to single bending magnet, it was planned to inject the beam into the undulators with 22.5 degrees bending, as shown in Fig. 1. In addition, the magnet material choice for the undulators may probably be SmCo or NbFe. On the other hand, single pass gain variation versus electron beam energy and K parameter for U25 (left) and U90 (right) undulators [3], are shown in Fig. 7.

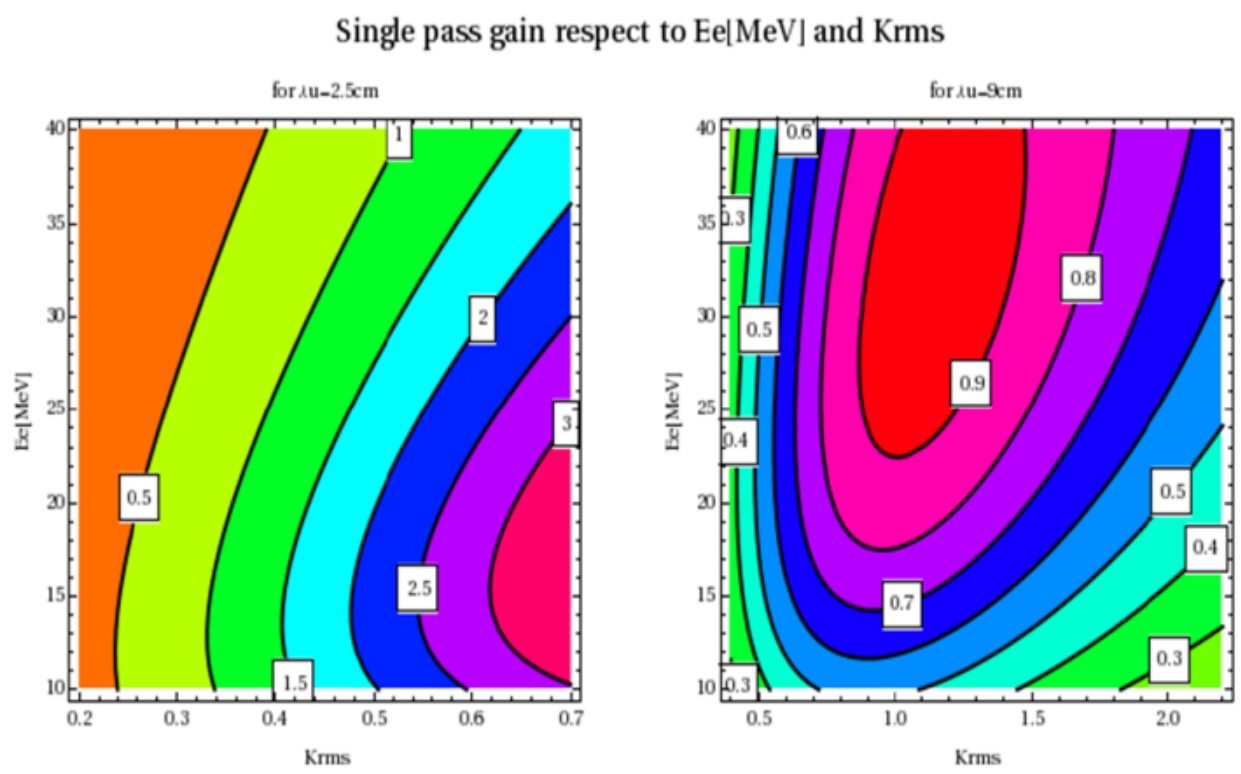

Figure 7: Single pass gain vs $E_e$(MeV) & K for U25 (left) and U90 (right) @ 120 pC bunch charge and 1 ps bunch length

The optical cavity and undulator parameters are given in Table 2. And also, expected wavelength range for U25 and U90 is shown in Fig. 8. It is possible to obtain FEL between 2.5 - 250 μm ranges without taking into account the nonlinear effects.

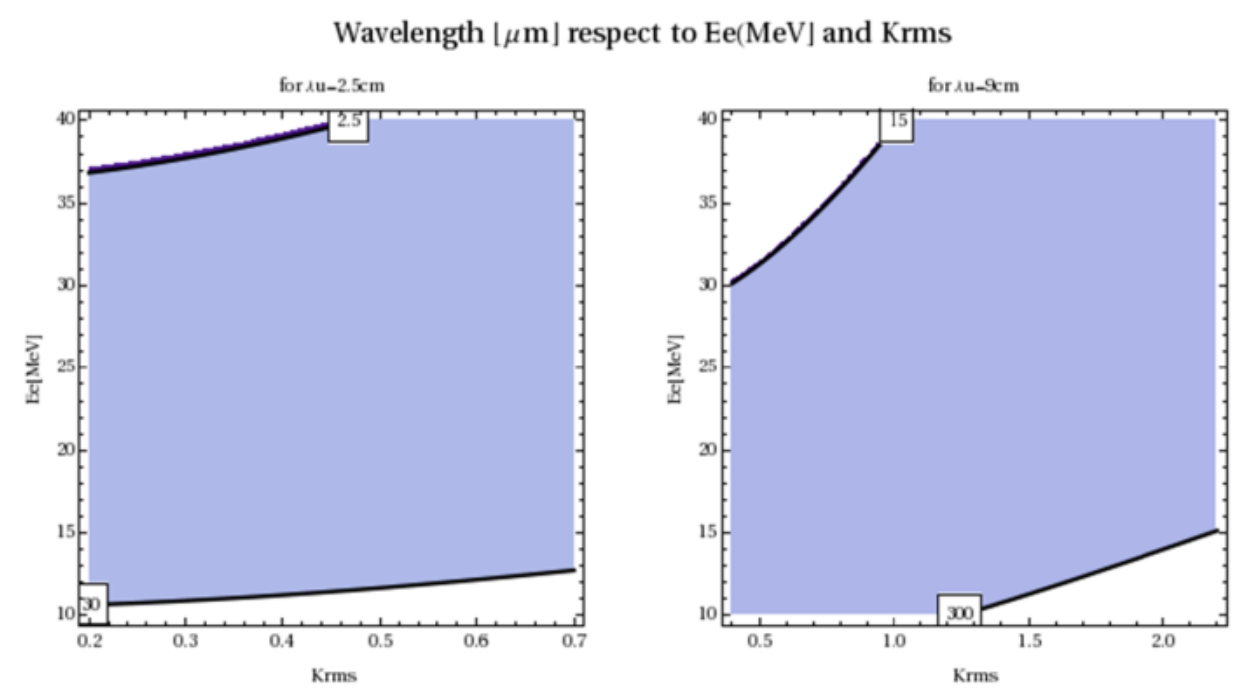

Figure 8: Obtainable wavelength ranges (μm) vs $E_e$(MeV) & K for U25 (left) and U90 (right)

FELO software [9], a one-dimensional free electron laser oscillator simulation code developed by ASTeC CCLRC Daresbury laboratory, was used for FEL optimizations.

Table 2: TAC IR FEL optical cavity and undulator parameters

| Parameter | Resonator-1 | Resonator-2 |
|---|---|---|
| Magnet material | SmCo | SmCo |
| Period length (mm) | 25 | 90 |
| Number of periods | 60 | 40 |
| Magnet dimensions (mm) | 74x26x10.5 | 90x90x35 |
| Steel pole dimensions (mm) | 74x18x2 | 70x20x10 |
| Magnetic gap (mm) | 15 | 40 |
| Effective field (T) | 0.3591 | 0.4205 |
| $K_{rms}$ | 0.71 | 2.5 |
| Optical cavity length (m) | 11.53 | 11.53 |
| Resonator type | Symm. conc. | Symm. conc. |
| 1st mirror, rad. of curv. (m) | 5.86 | 6.32 |
| 2nd mirror, rad. of curv. (m) | 5.86 | 6.32 |
| Rayleigh length, $Z_R$ (m) | 0.75 | 1.8 |
| Mirror material | Au / Cu | Au / Cu |
| Rad. of out coup. hole (mm) | 01 / 02 / 09 | 2 / 3 / 4 |
| Type of waveguide | Not determin. | Not determin. |

Some of the FEL optimizations are given in figures below. Firstly, FEL gain and power vs number of passes for seperate wavelengths obtained from U25 & U90 undulators, are shown in Fig. 9 and Fig.10 respectively.

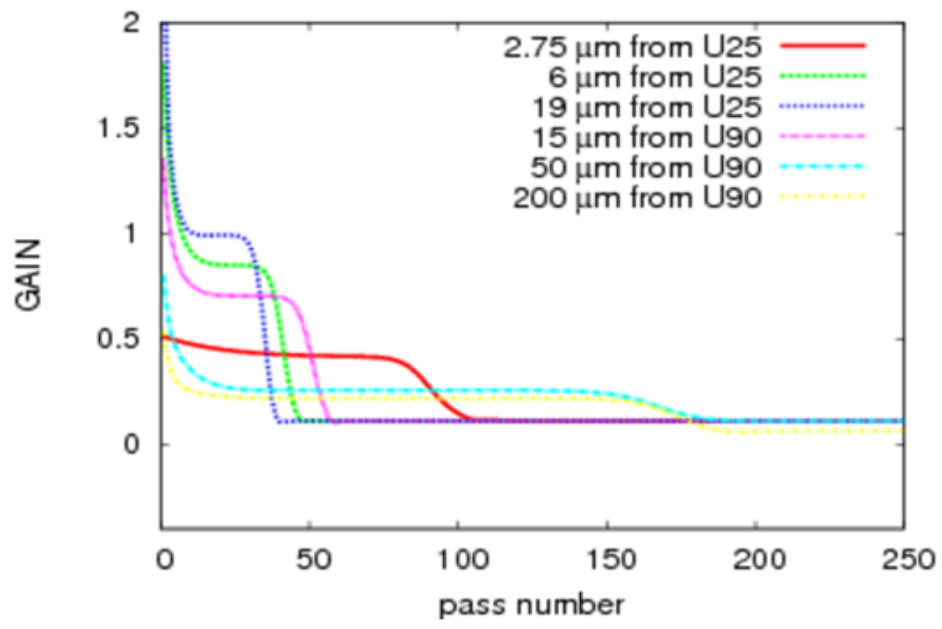

Figure 9: Gain vs number of passes

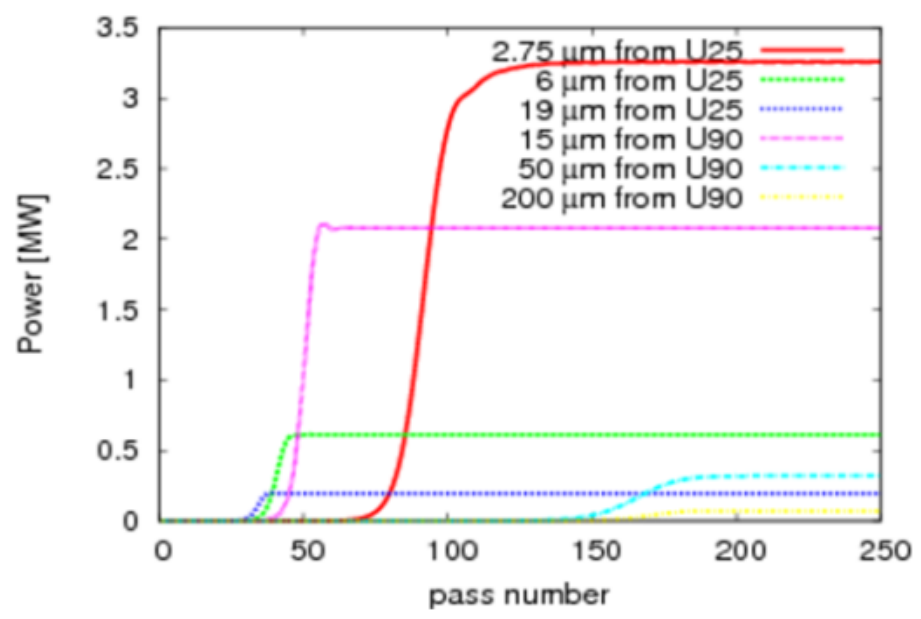

Figure 10: Intracavity power [MW] vs number of passes

Figure 11 shows laser pulse energy variation vs number of passes for seperate wavelengths obtained from U25 & U90 undulators.

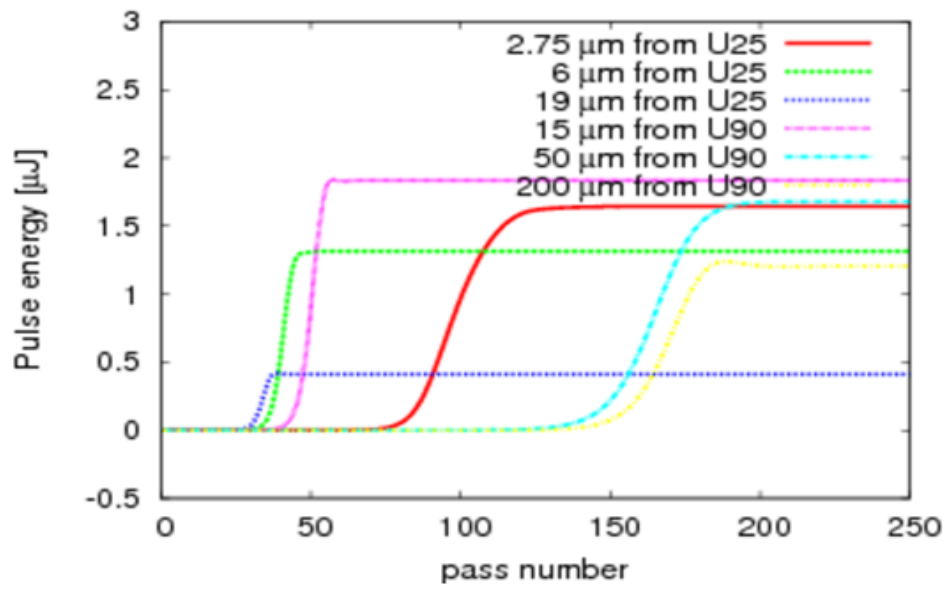

Figure 11: Pulse energy [μJ] vs number of passes

## CONCLUSION

The TAC IR FEL facility [1] will be the first accelerator based light source in Turkey. Thereby, Turkish scientists and researchers will become acquainted with accelerator technologies & light sources based on them. With no doubt, TAC IR FEL will provide new research opportunities in basic and applied sciences.

## ACKNOWLEDGEMENT

The authors would like to thank to all the contributors for this study on behalf of the TAC Project. And also, special thanks to the Turkish State Planning Organization (SPO) for supporting the TAC project and to the staff for their complete assistance. In addition, thanks a lot to ELBE staff, Forschungszentrum Dresden-Rossendorf [6].